\documentclass[conference]{IEEEtran}
\IEEEoverridecommandlockouts
\usepackage{amsmath,amssymb,amsfonts}
\usepackage{algorithmic}
\usepackage{graphicx}
\usepackage{textcomp}
\usepackage{xcolor}
\usepackage[section]{placeins}
\usepackage{url}
\usepackage{parskip}
\begin{document}
\title{ An Efficient Multi-threaded
Collaborative Filtering Approach in
Recommendation System}

\author{


\IEEEauthorblockN{Mahamudul Hasan}
\IEEEauthorblockA{
Department of Computer Science and Engineering\\
University of Minnesota Twin Cities\\
Minneapolis, United States \\
Email: munna09bd@gmail.com                 
}

}

\maketitle

\begin{abstract}
Recommender systems are a subset of information filtering systems designed to predict and suggest items that users may find interesting or relevant based on their preferences, behaviors, or interactions. By analyzing user data such as past activities, ratings, and preferences, these systems generate personalized recommendations for products, services, or content, with common applications including online retail, media streaming platforms, and social media. Recommender systems are typically categorized into three types: content-based filtering, which recommends items similar to those the user has shown interest in; collaborative filtering, which analyzes the preferences of similar users; and hybrid methods, which combine both approaches to improve accuracy. These systems enhance user experience by reducing information overload and providing personalized suggestions, thus increasing engagement and satisfaction. However, building a scalable recommendation system capable of handling numerous users efficiently is a significant challenge, particularly when considering both performance consistency and user data security, which are emerging research topics. The primary objective of this research is to address these challenges by reducing the processing time in recommendation systems. A multithreaded similarity approach is employed to achieve this, where users are divided into independent threads that run in parallel. This parallelization significantly reduces computation time compared to traditional methods, resulting in a faster, more efficient, and scalable recommendation system that ensures improved performance without compromising user data security.
\end{abstract}

\section{Introduction}
In this research study, we aim to advance the field of Recommendation Systems (RS) by introducing new methodologies and insights. Recommendation Systems have revolutionized how we access and interact with information, providing valuable recommendations and uncovering significant insights over recent years. Our work builds upon this foundation by adding a novel dimension to the understanding and functionality of RS.\\
At the core of RS is the concept of similarity, a fundamental aspect of human behavior. People naturally recommend things they enjoy to others, and this principle has driven the development of Recommendation Systems ~\cite{sappadla2017movie}. These systems filter and suggest information based on users' past preferences, leveraging a vast amount of data to generate meaningful recommendations. The definition and application of RS have evolved substantially over the past 14 years, with continuous advancements refining its scope and effectiveness ~\cite{bergh2023personalized}.\\
Recommendation Systems operate based on the principle of similarity ~\cite{kumar2015movie}. By analyzing the interactions and preferences of users, RS can make personalized recommendations to individuals with similar tastes. This process involves tracking user behavior and storing data, which is then used to optimize and structure recommendations effectively ~\cite{liu2014new}. The ability to analyze and interpret user data plays a crucial role in ensuring that recommendations are relevant and valuable.\\
RS can be broadly categorized into two main approaches: Content-Based Recommendation and Collaborative Recommendation. These approaches form the foundation of RS and each has its own methodology and application. Content-Based Recommendation Systems (CBRS) focus on analyzing user profiles to suggest content that aligns with individual preferences. CBRS identifies similarities between content and user interests to make recommendations. On the other hand, our research specifically investigates Collaborative Recommendation Systems (CRS). CRS aims to predict user ratings for items based on historical data and previously given reviews. By storing and analyzing a large volume of user reviews, CRS can provide personalized recommendations to new users. This method often employs memory-based techniques, where similarity calculations are used to identify and select active neighbors based on their shared preferences ~\cite{cheng2016proceedings}.\\
In addition to exploring different approaches in RS, our research also emphasizes the importance of efficient implementation techniques. Multi-threading is a critical aspect of program execution that significantly enhances performance. Traditional single-threaded programs execute one task at a time, but multi-threading allows multiple processes to run concurrently. This approach maximizes CPU utilization by enabling simultaneous execution of different threads within the same application. Multi-threading is particularly beneficial for programs that require parallel processing, as it reduces execution time and improves overall efficiency. By employing multi-threading, we have been able to minimize program execution time, achieving greater economic efficiency and aligning with our goal of reducing time consumption.\\

Overall, this research contributes to the field of Recommendation Systems by introducing new methodologies and enhancing existing approaches. By leveraging the principles of similarity and employing advanced implementation techniques like multi-threading, we aim to advance the effectiveness and efficiency of Recommendation Systems.

\section{Literature Review}
\label{Literature Review}
In the pursuit of addressing scalability issues within recommender systems (RS), a review of significant research background is crucial for a comprehensive understanding. Recommender systems, especially those utilizing collaborative filtering (CF), are extensively recognized for their utility. Multi-threading, a widely adopted technique for parallel processing, plays a pivotal role in reducing computational time. Traditional CF methods, despite their popularity, often struggle with the challenge of recommending suitable movies as the volume of information grows. To mitigate this challenge, several clustering algorithms have been proposed, including K-means, BIRCH, Mini-Batch K-means, Mean-Shift, Affinity Propagation, Agglomerative Clustering, and Spectral Clustering. Among these, K-means has demonstrated superior performance in addressing scalability and data sparsity issues ~\cite{cintia2020design}.\\

In the realm of individual health and fitness tracking, applications such as Apple Watch, Google Fit, and Samsung Health have become commonplace. These platforms continuously evolve to handle contextual information and nutritional data more effectively. Hybrid algorithms have been employed to tackle cold start and scalability challenges in health services. Parameters such as K, U, FM, FOOD, ID, si, square, and preference matrices are utilized in these algorithms to predict user similarities, hybrid learning predictions, and dynamic contexts. The accuracy improvement achieved through these hybrid methods is approximately 14.61
The rise of e-commerce has resulted in the processing of vast amounts of data, necessitating advanced techniques for accurate recommendations. Deep learning algorithms have been explored to enhance performance, including Multilayer Perceptrons with Auto-Encoders, Convolutional Neural Networks (CNN), Restricted Boltzmann Machines (RBM), Adversarial Networks (AN), and Neural Autoregressive Distribution Estimation. In rating-based systems, where ratings range from 1 to 5, and datasets encompass 10,000 patients across 500 hospitals, a combined RBM-CNN approach has shown the best results. This method, central to this research, has been evaluated using RMSE and MAE metrics ~\cite{sahoo2019deepreco}.\\
A study focused on scalability and efficiency within recommender systems highlights the use of neural networks to encode, compress, and fuse various contexts. Scalable systems are crucial as they accommodate new users without impacting existing data, allowing for flexible and efficient updates. This research addresses issues in Latent Factor Analysis (LFA) where parallelism is disabled due to dependencies between user and item latent features during training epochs. The proposed DASGD approach aims to resolve these challenges ~\cite{jiang2020factorization}.\\
The cold start problem, common in recommender systems, has been addressed through niche approaches that also tackle density issues. The JAC approach has been employed across various experiments to assess similarity ~\cite{zhang2019addressing}. Optimal values for K and basis size have been determined, with performance gains reaching 81.63\%. The system described in ~\cite{sarwar2002incremental} handles large volumes of data effectively, offering highly personalized recommendations through nearest neighbors and latent factors ~\cite{zhang2016grorec}.\\
The efficiency of modern movie services has increased, with platforms integrating social networking services (SNS) and location-based SNS (LBSNS). Mobile Personalized Recommendations (MPR) provide users with relevant information while mitigating information overload. Multithreading enhances performance, achieving speeds 5 to 7 times faster than traditional methods. The Slope-1 algorithm updates data efficiently, and comparisons with sequential algorithms show a speed increase of 9 times using 16 threads ~\cite{karydi2012multithreaded}.\\
The exponential growth of digital documents on the web presents a challenge in organizing content according to user preferences. Support Vector Machines (SVM) and CUDA GPS parallelism have been utilized to improve this process, with SVM achieving an accuracy of approximately 93.4\% and 8\% CPU usage ~\cite{chatterjee2019text}.\\
For predicting user ratings, parallel graph matrices are employed to identify and compute missing ratings. Recommender systems, prevalent in e-commerce platforms such as Facebook and Netflix, benefit from the parallel matrix graph approach for enhanced scalability and speed. KNN is used for similarity measurement ~\cite{koohi2019parallel}, comparing favorably with sequential methods.\\
Apache Spark, an open-source clustering framework, facilitates fault tolerance and parallel data processing. It operates with a master node and multiple slave nodes, enabling efficient execution and management of tasks. Spark supports both horizontal and vertical scaling, optimizing performance for various data processing needs ~\cite{jiang2019parallel} ~\cite{wang2020parallelization}.

\section{Methodology}
\label{Methodology}
Recommender systems are a specialized subset of information filtering systems, designed to predict and suggest items that users may find relevant or interesting based on their preferences, behaviors, or interactions. By analyzing user data such as past activities, ratings, and preferences, these systems generate personalized recommendations for products, services, or content. Common applications include online retail, media streaming platforms, and social media, where tailored suggestions enhance user satisfaction. Recommender systems can be categorized into three main types: content-based filtering, which recommends items similar to those a user has shown interest in; collaborative filtering, which analyzes the preferences of similar users to make suggestions; and hybrid methods, which combine both approaches to improve the accuracy of the recommendations.\\

Despite their growing popularity, building a scalable recommendation system that can efficiently handle large numbers of users poses significant challenges. These include ensuring both performance consistency and the security of user data, which are key areas of emerging research. Personalized recommendations not only need to be accurate but must also be generated quickly to keep users engaged, especially in high-traffic environments. The objective of this research is to address these challenges by focusing on reducing the processing time in recommendation systems, without compromising data security.\\

The proposed method introduces a multithreaded similarity approach, where users are divided into independent threads that run in parallel. By parallelizing the computations, the system can process large amounts of data more efficiently, significantly reducing computation time compared to traditional methods. This results in a faster, more scalable recommendation system that enhances performance and maintains user data security. Evaluation metrics based on the reduced processing time have been extracted to assess the method's effectiveness, as depicted in Figure 1, demonstrating the potential of this approach to overcome the limitations of conventional recommendation systems.

\begin{figure}[hbt!]
    \centerline{\includegraphics[width=1\linewidth]{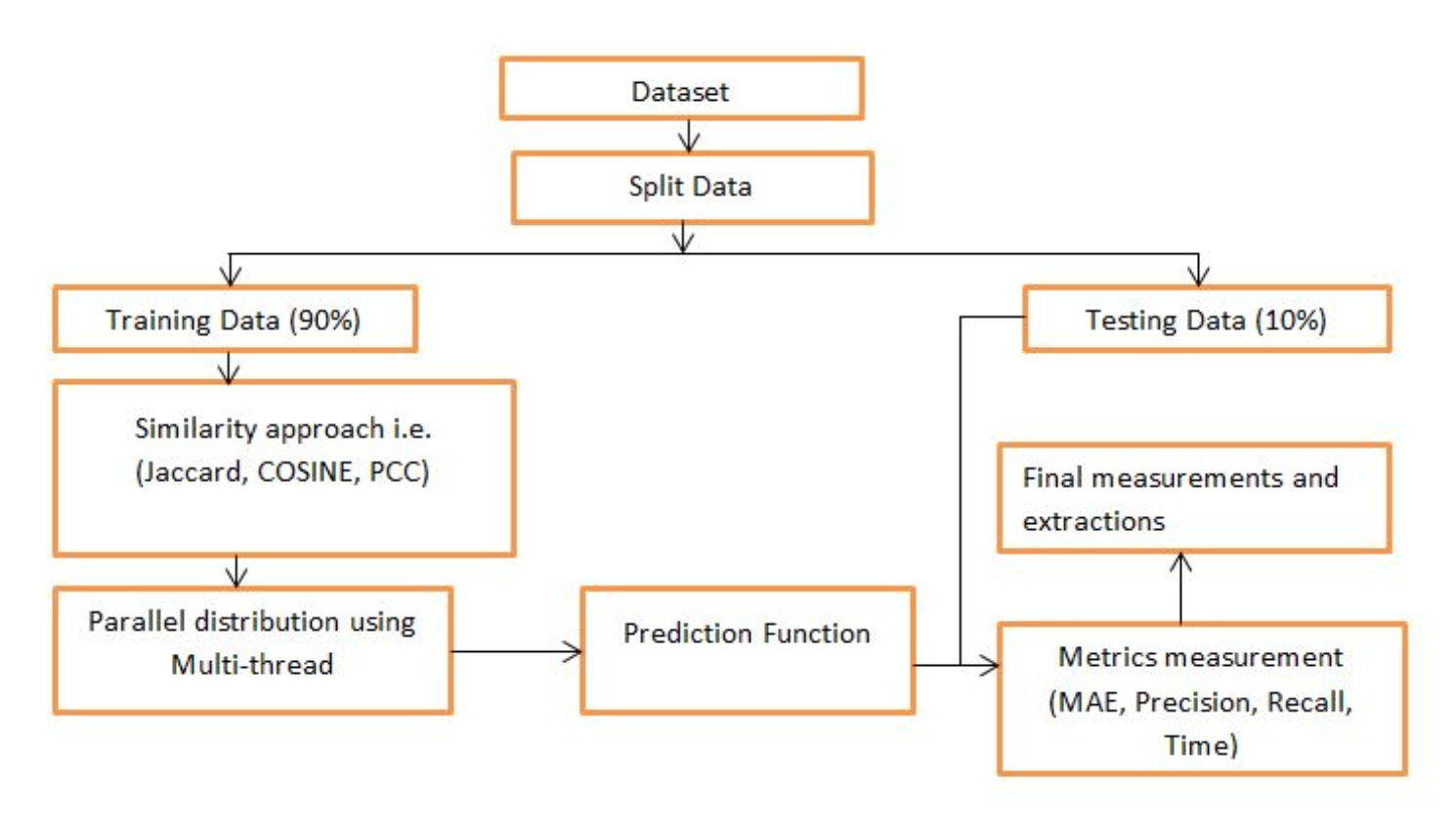}}
    \caption{Proposed Method}
\label{activity}
\end{figure}

\section{Similarity Approaches}
\subsection{Jaccard Similarity}
Jaccard is an approach that is applied to find the rating-based similarity from the given rating list. It is used to find similarities and dissimilarities of datasets by comparing them. In Jaccard, two individual users’ ratings are compared. It is applied only when the same items are rated by both users. Jaccard measures the similarity in a division module where the common features are divided by all given properties. The similarity between two users has been expressed below as Equation 1 and depicted as Jaccard similarity -

\begin{equation}
similarity(m, n)Jaccard = \frac{|p_m| \cap |p_n|}{|p_m| \cup |p_n|}
\end{equation}

\subsection{Pearson Correlation Coefficient - (PCC) Similarity}
Previous approaches have explored user similarity by comparing ratings and absolute values, but the Pearson Correlation Coefficient (PCC) introduces a more advanced approach by examining the linear relationship between two vectors \cite{agarwal2017similarity}. Since PCC is a statistical measure, its values range from -1 to 1. To maintain consistency across all similarity measures, we normalize the PCC values to fall between 0 and 1. The mathematical model of PCC is presented in Equation 2.

\begin{equation}
pcc =\frac{ \sum_{q\epsilon I}(r_m,_q-\bar{r_m})(r_n,_q-\bar{r_n}) }{%
\sqrt{\sum_{q\epsilon I}(r_m,_q-\bar{r_m})^2}\sqrt{\sum_{q\epsilon I}(r_n,_q-\bar{r_n})^2}}
\end{equation}

\section{Evaluation Metrics}
\subsection{MAE}
The Mean Absolute Error (MAE) measures the distance between predicted and actual values \cite{basilico2004unifying}. It is calculated by comparing the predicted values with the actual values in the dataset, taking the absolute difference, and then averaging these errors \cite{sarwar2001item}. A lower MAE value indicates better model performance. The mathematical expression for MAE is provided below.
\begin{equation}
MAE = \frac{ \sum_{n = 1}^{Mu} |p_u,_n - r_u,_n| }{N_u}
\end{equation}
where ru,n is the actual rating pu,n is the predicted rating, and Mu is the total rating of
user u.

\subsection{Precision}
Precision is the measurement of accuracy that counts the percentage of pertinent items from all the items that have been retrieved ~\cite{isinkaye2015recommendation}. For better performance, the measure should be as high as possible. The computed formula is –

\begin{equation}
Precision = \frac{TP}{TP+FP}
\end{equation}
Here, true positive ratings are those that were predicted as good and the actual rating was also that. The ratings of positives and negatives were counted within a threshold. The value of both ratings needed to be measured in precision.

\subsection{Recall}
Recall stands for the complete metric that holds the percentage of the retrieved items excerpted from all the pertinent items. The formula here is illustrated below –

\begin{equation}
Recall = \frac{TP}{TP+FN}
\end{equation}
False-negative is the value that is actually rated in the good range but predicted as bad.

\subsection{F-Score}
The F-Score metric is the comparing harmonic mean of Precision and Recall. As precision and recall evaluations do not provide significance, the f-score invents the balance and combines both evaluations for the best performance.

\begin{equation}
F1 = \frac{2*Precision*Recall}{Precision+Recall}
\end{equation}

\section{Empirical Analysis}
\subsection{Dataset}
The dataset is a crucial component of any research paper. In this study, we utilized the publicly available MovieLens 1M dataset \cite{khan2020enriching}, which contains 4,000 movie ratings from 6,040 users. The dataset provides millions of anonymous ratings, making it highly suitable for research on Collaborative Filtering-based Recommender Systems \cite{shi2008personalized}. For our experiments, we split the data, allocating 90\% for training and 10\% for testing. This division ensures a robust evaluation of the model's performance.

\subsection{Time Comparison}
The key aspect of the results involves applying a multi-threaded approach to traditional algorithms. We implemented the Jaccard, Cosine, and Pearson Correlation similarity measures individually to establish baseline processing times. By distributing these algorithms across multiple threads, we achieved a reduction in overall processing time. Figure 1 illustrates the time comparison between the traditional and multi-threaded approaches.

\subsection{MAE Graph}
MAE (Mean Absolute Error) evaluates the minimum error among the top neighbors chosen in a user-based Collaborative Filtering (CF) approach. Neighborhood selection is a critical step in CF, where top neighbors are selected from existing candidates, including active users. Figure 2 illustrates the MAE values, showing that initial errors are relatively high but decrease as the amount of data increases. This trend indicates that with a sufficient amount of data, the error associated with the top neighbors improves.

\begin{figure}[hbt!]
    \centerline{\includegraphics[width=1\linewidth]{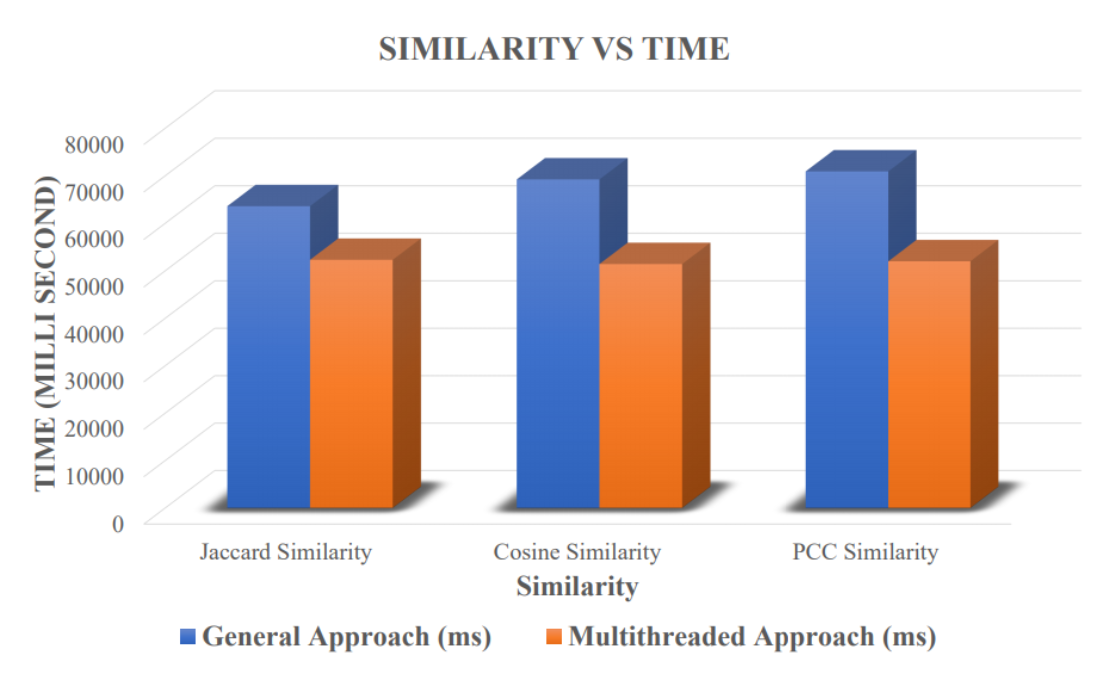}}
    \caption{Time Comparison}
\label{activity2}
\end{figure}
amount of data, MAE ~\cite{natarajan2020resolving} became constant at one point and values will be unchanged in the future if data has been increased.

\subsection{Top-N and Sparsity vs Precision}
The second evaluation metric is Precision, which measures the accuracy of predicted relevant ratings compared to the original data. Precision is assessed based on the selection of top neighbors. Unlike MAE, which showed a decrease in error with more data, Precision increases as the amount of data grows \cite{ahmadian2020social}. In other words, Precision improves with the increase in data volume. Figure 3 illustrates the relationship between the top-N neighbors and Precision, demonstrating how Precision rises with more data.

\subsection{Top-N vs Recall}
Recall measures how well the predicted ratings match the actual ratings, reflecting the ability to replicate the actual data based on predictions. The behavior of Recall is similar to that of Precision: as the amount of data increases, Recall also improves, enhancing overall accuracy. The graph initially shows a noticeable improvement in Recall with increasing data, but as more data is added, the rate of increase in Recall stabilizes. Figure 4 illustrates the relationship between top-N neighbors and Recall, highlighting the trend in performance as data volume grows.

\subsection{Top-N vs F-Score}
The F-score combines both Precision and Recall into a single metric, providing a comprehensive evaluation of model performance. Figure 5 illustrates the F-score, showing how it integrates Precision and Recall to offer a balanced measure of accuracy.

\begin{figure}[hbt!]
    \centerline{\includegraphics[width=1\linewidth]{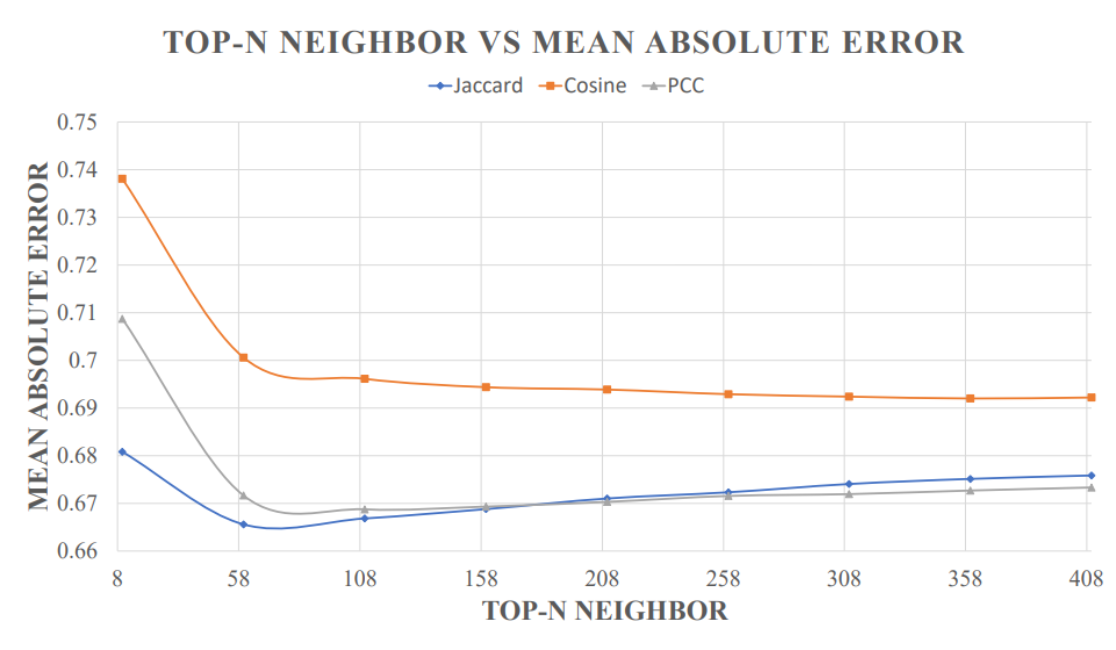}}
    \caption{Top-N vs MAE}
\label{activity3}
\end{figure}

\begin{figure}[hbt!]
    \centerline{\includegraphics[width=1\linewidth]{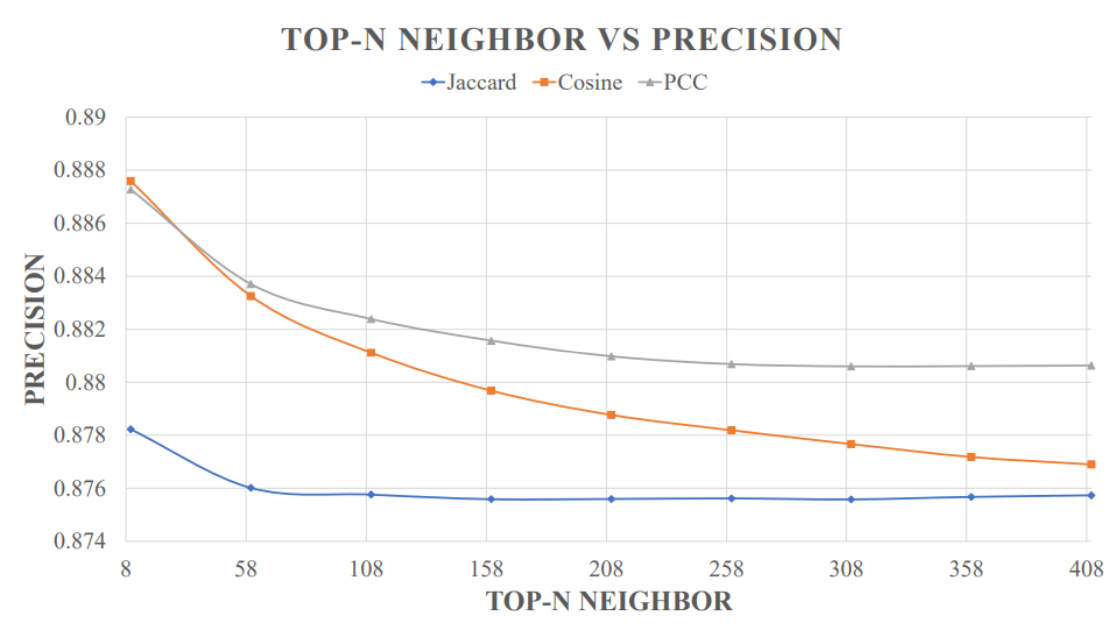}}
    \caption{Top-N vs Precision}
\label{activity4}
\end{figure}

\begin{figure}[ht]
    \centerline{\includegraphics[width=1\linewidth]{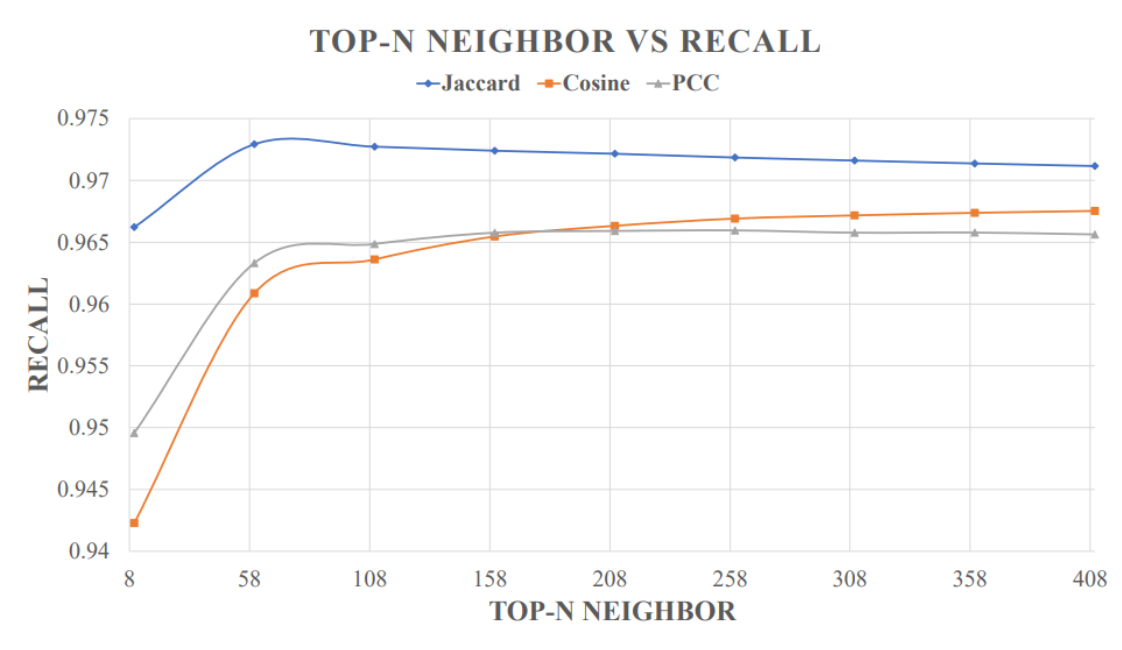}}
    \caption{Top-N vs Recall}
    \label{activity5}
\end{figure}

\begin{figure}[ht]
    \centerline{\includegraphics[width=1\linewidth]{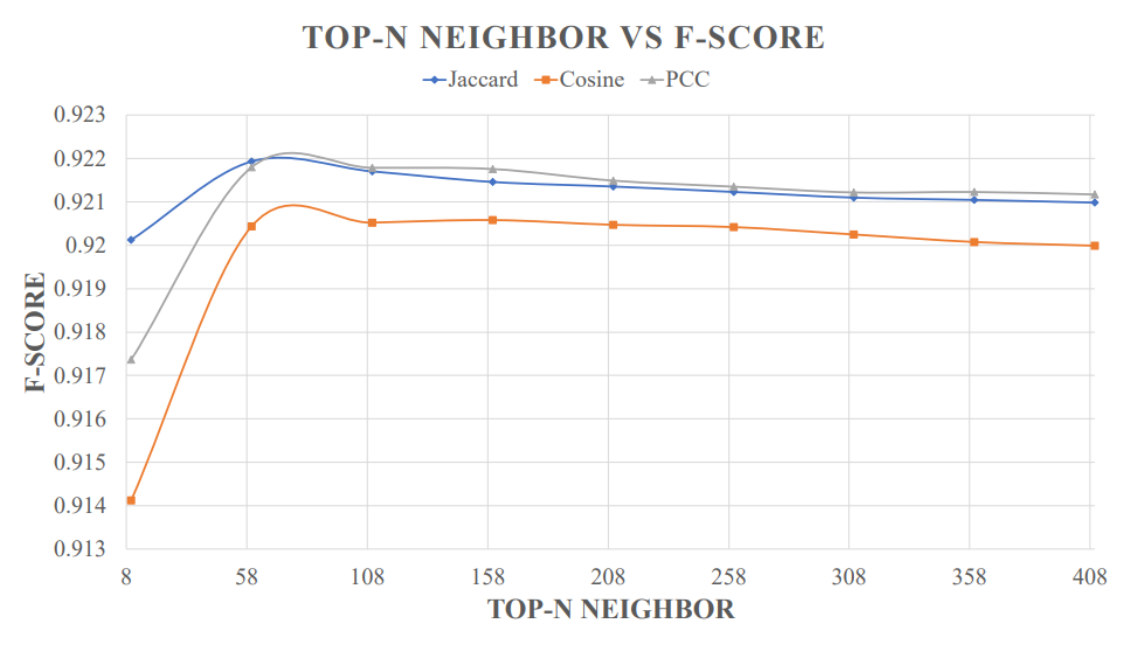}}
    \caption{Top-N vs F-Score}
    \label{activity6}
\end{figure}

\section{Conclusion}
\label{Conclusion}

Recommendation Systems (RS) have become integral across various fields, including machine learning, big data, image processing, and data science. Recently, scalability has emerged as a significant challenge in RS, and researchers have proposed numerous solutions to improve upon existing methods. While state-of-the-art algorithms are widely used in RS, our work takes a novel approach by utilizing multi-threading to reduce CPU processing time. Our research demonstrates a substantial reduction in processing time compared to conventional methods, effectively addressing scalability issues. The results, as illustrated through comparative figures, clearly show that processing time has been significantly optimized.\\

Time efficiency is a critical factor in research. Prolonged training and processing times are not ideal indicators of robust research. In this study, we addressed the issue by distributing the algorithm's tasks into smaller portions and integrating them efficiently, leading to improved performance in less time. The multi-threading approach played a key role in minimizing execution time and enhancing the overall cost-effectiveness of the program. We employed Jaccard, Cosine, and Pearson Correlation Coefficient (PCC) methods to determine similarity, which proved to be a successful innovation in reducing both time and computational complexity.\\

While we achieved our primary objectives, there is still room for further improvement in the system's core modules. In the future, we aim to extend this work by incorporating data from various social media platforms, including images, videos, check-ins, and comments, for deeper social media analysis. Additionally, by utilizing geolocation data, we could predict crime hotspots and study behavioral patterns to prevent crimes before they occur.

\bibliographystyle{IEEEtran}
\bibliography{refernce}
\end{document}